\documentclass[aps,PRB,twocolumn,superscriptaddress,preprintnumbers]{revtex4-2}
\usepackage[T1]{fontenc}
\usepackage{times}
\usepackage{graphicx,color}
\usepackage{amsfonts,amsmath,amssymb,amsbsy}
\usepackage[colorlinks=true, linkcolor=blue, citecolor=blue, urlcolor=blue]{hyperref}
\usepackage{float}

\begin{document}

\title{Nonlinear spin-Seebeck diode in $f$-wave magnets, third-order
spin-Nernst effects in $g$-wave magnets and spin-Nernst effects in $i$-wave
altermagnets}
\author{Motohiko Ezawa}
\affiliation{Department of Applied Physics, The University of Tokyo, 7-3-1 Hongo, Tokyo
113-8656, Japan}

\begin{abstract}
A prominent feature of $d$-wave altermagnets is that spin current is
generated by applying temperature gradient, which is known as the
spin-Nernst effect. We show in $f$-wave magnets that spin current is
generated proportional to the square of the temperature gradient, which we
call the nonlinear spin-Seebeck current. It can be used as a spin current
diode. In addition, we show in $g$-wave altermagnets that spin current is
generated in the third order of the temperature gradient. We also show in $i$%
-wave altermagnets that spin current is generated perpendicular to the
temperature gradient, which is the spin-Nernst current. We have derived
analytic formulas for these spin currents. It is interesting that these
phenomena occur in the absence of the spin-orbit interaction. On the other
hand, we show in $p$-wave magnets that spin current is not generated by
temperature gradient.
\end{abstract}

\date{\today }
\maketitle

\textbf{Introduction:} Spin current conveys spins without net charge
current. Spin current generation is a key concept in spintronics. It is
generated by the spin-Hall effect perpendicular to applied electric field in
the presence of spin-orbit interactions\cite{Dya,Dya2,Sinova}. In addition,
spin current is generated by applying temperature gradient $\nabla T$, which
is known as the spin-Seebeck effect\cite{Uchida,Bauer}. In addition, spin
current flows perpendicular to temperature gradient, which is known as the
spin-Nernst effect\cite{SqChen,Meyer}. These are linear responses. On the
other hand, there are several studies on nonlinear responses by applying
electric field\cite%
{Gao,Sodeman,Ideue,HLiu,Michishita,Watanabe,CWang,Oiwa,AGao,NWang,KamalDas,Kaplan,Ohmic,Xiang,ZGong,EzawaMetricC,YFang,EzawaPNeel}%
. There are also a few studies on nonlinear responses to temperature gradient%
\cite{Yu,Karki,Zeng,March,Arisawa,Hirata,Vars,Vars2,YFZ}. Nonlinear spin
current generation\cite{Hamamoto,Kameda,Hayami22B,Hayami24B,GI} is
interesting for spintronics applications. Especially, the second-order spin
current generation can be used as a spin current diode\cite{GI}. It is an
intriguing problem to search nonlinear spin current generations induced by
temperature gradient such as the nonlinear spin-Seebeck effect and the
nonlinear spin-Nernst effect.

Altermagnets attract much attention in the context of spintronics\cite%
{SmejX,SmejX2}, which include $d$-wave, $g$-wave and $i$-wave altermagnets
in the view point of the crystal symmetry. In addition, there are $p$-wave%
\cite{Hayami,pwave,He} and $f$-wave magnets\cite{pwave,He}. They are
antiferromagnets and have spin-split band structures characterized by the
number of nodes: See Table \ref{TableA}. A prominent feature of $d$-wave
aletermagnets is that spin current is generated perpendicular to electric
field without using spin-orbit interactions\cite%
{Naka,Gonza,NakaB,Bose,NakaRev}. It is also known that spin current is
generated perpendicular to temperature gradient in $d$-wave altermagnets,
which is the spin-Nernst effect\cite{Naka,APEX}. It is known\cite{GI} that
spin current is generated proportional to the square of applied electric
field in $f$-wave magnets, which can be used for a perfect spin current
diode. It is also known that the third-order and the fifth-order nonlinear
spin currents are generated in $g$-wave and $i$-wave altermagnets,
respectively.

In this paper, we investigate nonlinear spin current generation by
temperature gradient in two dimensions. We derive the formula for the
current generation as nonlinear responses to temperature gradient based on
the Boltzmann equation valid up to any order in temperature gradient. Then,
we show in $f$-wave magnets that spin current is generated parallel to the
square of the temperature gradient $\left( \nabla T\right) ^{2}$ implying a
nonreciprocity. It can be used as a spin current diode, which we call the
nonlinear spin-Seebeck diode. It is interesting that this phenomenon occurs
in the absence of spin-orbit interactions. In addition, linear spin current
is generated perpendicular to temperature gradient in $i$-wave altermagnets.
It is contrasted to spin current induced by electric field, where there is
no linear response. Furthermore, we show in $g$-wave altermagnets that there
is the third-order response to temperature gradient. Finally, we show that
no linear nor nonlinear spin current is generated in $p$-wave magnets. They
are summarized in Table \ref{TableA}. 
\begin{table}[h]
\begin{tabular}{|c|c|c|c|c|c|c|}
\hline
& $s$ & $p$ & $d$ & $f$ & $g$ & $i$ \\ \hline
nodes & 0 & 1 & 2 & 3 & 4 & 6 \\ \hline
Mirror $M_{y}$ & Yes & Yes & No & Yes & No & No \\ \hline
1st and 3rd spin-Nernst & No & No & $j_{\text{spin}}^{(x;y)}$ & No & $j_{%
\text{spin}}^{(x^{3};y)}$ & $j_{\text{spin}}^{(x;y)}$ \\ \hline
Mirror $M_{x}$ & Yes & No & No & No & No & No \\ \hline
2nd spin-Seebeck & No & No & No & $j_{\text{spin}}^{(x^{2};x)}$ & No & No \\ 
\hline
\end{tabular}%
\caption{Fermi-surface symmetry and current induced by temperature gradient
in two dimensions. It is found that the second-order spin-Seebeck effect
occurs only in $f$-wave magnets.}
\label{TableA}
\end{table}

\textbf{Nonlinear spin-Seebeck and spin-Nernst effects:} We derive a current
induced by temperature gradient $\mathbf{\nabla }T$ based on the Boltzmann
equation. By assuming the spatial dependence of temperature $T\left( \mathbf{%
r}\right) $, the Fermi distribution function is given by $f(\mathbf{r},%
\mathbf{k})=1/\left( \exp \left( \left( \varepsilon \left( \mathbf{k}\right)
-\mu \right) /\left( k_{\text{B}}T\left( \mathbf{r}\right) \right) \right)
+1\right) $. The Boltzmann equation is given by%
\begin{equation}
\frac{df(\mathbf{r},\mathbf{k})}{dt}=-\frac{f(\mathbf{r},\mathbf{k}%
)-f^{\left( 0\right) }}{\tau },
\end{equation}%
where $\tau $ is the relaxation time.

We assume that the temperature gradient is along the $a$ direction and we
measure the induced current along the $b$ direction. We expand the Fermi
distribution in powers of $\partial _{a}T$, $f=f^{\left( 0\right)
}+f^{\left( 1\right) }+\cdots $. The current is correspondingly expanded as $%
j=j^{(a^{0};b)}+j^{(a^{1};b)}+\cdots $, where%
\begin{equation}
j^{(a^{\ell };b)}=-e\int d^{D}kf_{a}^{\left( \ell \right) }v_{b}
\end{equation}%
with the velocity $v_{b}=\partial \varepsilon /\partial \left( \hbar
k_{b}\right) $ and dimension $D$. It is straightforward to solve the
Boltzmann equation recursively to derive the formula%
\begin{align}
j^{(x^{\ell };b)}=& -e\int d^{D}kf^{\left( \ell \right) }v_{b}  \notag \\
=& e\left( -\frac{\tau }{\hbar }\partial _{x}T\right) ^{\ell }\int
d^{D}k\left( \frac{\partial \varepsilon }{\partial k_{x}}\right) ^{\ell }%
\frac{\partial \varepsilon }{\partial k_{b}}\frac{\partial ^{\ell }f^{\left(
0\right) }}{\partial T^{\ell }},  \label{jCurrent}
\end{align}%
where we have assumed that temperature gradient is along the $x$-axis ($a=x$%
). See Supplementary Material I with respect to this formula.

At high temperature $T$, the derivative of the Fermi-distribution function
is approximated as%
\begin{equation}
\frac{\partial ^{\ell }f^{\left( \ell \right) }}{\partial T^{\ell }}=\frac{%
\ell !}{4}\beta ^{\ell +1}\left( \varepsilon -\mu \right) ,
\end{equation}%
where $\beta =1/k_{\text{B}}T$.

We consider a Hamiltonian which is diagonal with respect to the spin, $H=H_{%
\text{kine}}\sigma _{0}+H_{J}\sigma _{z}$, whose energy is given by $%
\varepsilon =\varepsilon _{\text{kine}}+s\varepsilon _{J}$ with $s=\pm 1$,
representing the spin up and down. We define the $\ell $-th order nonlinear
charge current by 
\begin{equation}
j_{\text{charge}}^{(x^{\ell };b)}=j_{\uparrow }^{(x^{\ell
};b)}+j_{\downarrow }^{(x^{\ell };b)},
\end{equation}%
while the $\ell $-th order nonlinear spin current by 
\begin{equation}
j_{\text{spin}}^{(x^{\ell };b)}=\frac{1}{2}\left( j_{\uparrow }^{(x^{\ell
};b)}-j_{\downarrow }^{(x^{\ell };b)}\right) ,
\end{equation}%
where $b=x,y$. The diagonal charge (spin) current $j_{\text{charge}%
}^{(x^{\ell };x)}$\ ($j_{\text{spin}}^{(x^{\ell };x)}$) induced by the $\ell 
$-th power of the temperature gradient $\left( \partial _{x}T\right) ^{\ell }
$\ is called the $\ell $-th Seebeck (spin-Seebeck) current. On the other
hand, off-diagonal charge (spin) current $j_{\text{charge}}^{(x^{\ell };y)}$%
\ ($j_{\text{spin}}^{(x^{\ell };y)}$) induced by the $\ell $-th power of the
temperature gradient $\left( \partial _{x}T\right) ^{\ell }$\ is called the $%
\ell $-th Nernst (spin-Nernst) current. There is such a case as a
ferromagnet where both charge and spin currents flow induced by the
temperature gradient $\partial _{x}T$,\ which is called the spin-polarized
Seebeck current.

\textbf{Symmetry analysis:} We first study the nonlinear spin-Seebeck
current. If there is a mirror symmetry $x\rightarrow -x$ in the energy, $%
\varepsilon \left( -k_{x}\right) =\varepsilon \left( k_{x}\right) $, we have%
\begin{equation}
\frac{\partial \varepsilon \left( -k_{x}\right) }{\partial k_{x}}=-\frac{%
\partial \varepsilon \left( k_{x}\right) }{\partial k_{x}}.
\end{equation}%
Then, it follows from (\ref{jCurrent}) that $j_{s}^{\left( x^{\ell
};x\right) }=0$ for even $\ell $ from Eq.(\ref{jCurrent}).

Next, we study the nonlinear spin-Nernst current. If there is a mirror
symmetry $y\rightarrow -y$ in the energy, $\varepsilon \left( -k_{y}\right)
=\varepsilon \left( k_{y}\right) $, we have%
\begin{equation}
\frac{\partial \varepsilon \left( -k_{y}\right) }{\partial k_{x}}=-\frac{%
\partial \varepsilon \left( k_{y}\right) }{\partial k_{x}}.
\end{equation}%
Then, the current is zero, $j_{s}^{\left( x^{\ell };y\right) }=0$. Hence,
there is no spin-Nernst effects in $f$-wave magnets in two dimensions.

Even if there is no mirror symmetry $y\rightarrow -y$ in the energy, if
there is a mirror symmetry $x\rightarrow -x$ in the energy, it follows from (%
\ref{jCurrent}) that $j_{s}^{\left( x^{\ell };x\right) }=0$ for odd $\ell $. 
\begin{figure}[t]
\centerline{\includegraphics[width=0.48\textwidth]{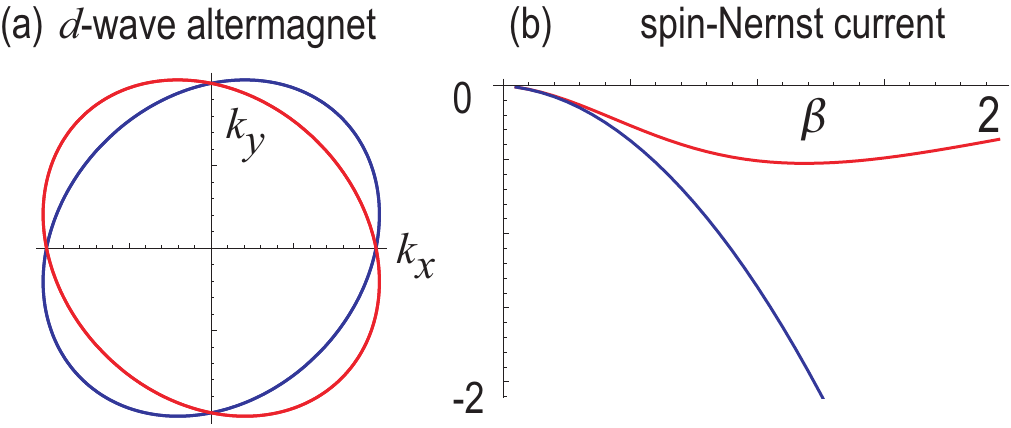}}
\caption{$d$-wave altermagnet. (a) Fermi surfaces with two nodes. Red oval
indicates the Fermi surface with up spin and blue oval indicates down spin.
(b) Spin current $j_{\text{spin}}^{\left( x;y\right) }/\partial _{x}T$ in
units of $e\protect\tau /\hbar $ as a function of the inverse temperature $%
\protect\beta =1/k_{\text{B}}T$. Red curve is a numerical result, while blue
curve is an analytic result base on the high-temperature expansion. We have
set $m=1$, $J=1/2$ and $a=1$.}
\label{FigD}
\end{figure}

$d$\textbf{-wave altermagnet}: We revisit the spin-Nernst effect in $d$-wave
altermagnets\cite{Naka,APEX} in the present formalism. The $d$-wave symmetry
is given by 
\begin{equation}
f_{d}^{2\text{D}}\left( \mathbf{k}\right) =k_{x}k_{y}.
\end{equation}%
The Fermi surface is shown in Fig.\ref{FigD}(a). The Seebeck current is
calculated as

\begin{equation}
\frac{j_{\text{charge}}^{\left( x;x\right) }}{\frac{e\tau }{\hbar }\partial
_{x}T}=-\frac{\pi ^{2}\beta ^{2}\left( 2+J^{2}m^{2}\right) \left( 2-m\mu
\right) }{4m^{3}}.
\end{equation}%
The Spin-Nernst current is calculated as

\begin{equation}
\frac{j_{\text{spin}}^{\left( x;y\right) }}{\frac{e\tau }{\hbar }\partial
_{x}T}=-\frac{\beta ^{2}J\pi ^{2}}{4m^{2}}.
\end{equation}%
It is shown in Fig.\ref{FigD}(b) as well as numerically obtained result
without using the high-temperature expansion.

\begin{figure}[t]
\centerline{\includegraphics[width=0.48\textwidth]{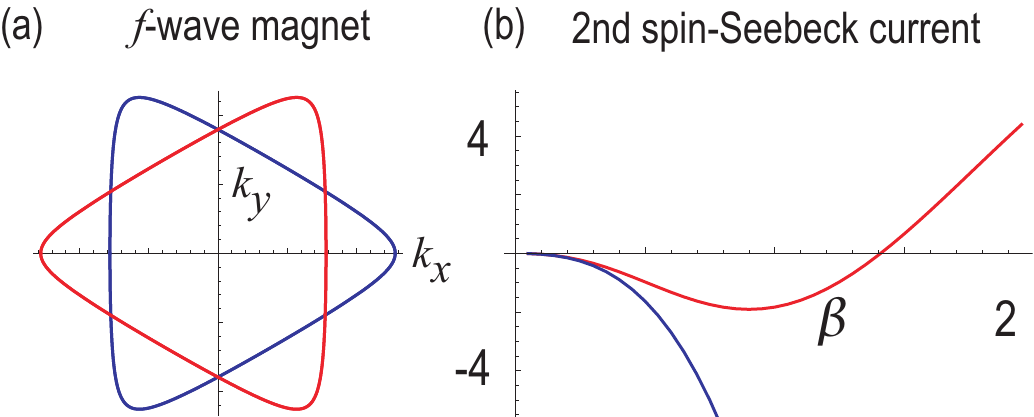}}
\caption{$f$-wave magnet. (a) Fermi surfaces with three nodes. (b) Spin
current $j_{\text{spin}}^{\left( x^{2};y\right) }/\left( \partial
_{x}T\right) ^{2}$ in units of $e\left( \protect\tau /\hbar \right) ^{2}$ as
a function of the inverse temperature $\protect\beta =1/k_{\text{B}}T$. See
also the caption of Fig.\protect\ref{FigD}.}
\label{FigF}
\end{figure}

$f$\textbf{-wave magnet:} The $f$-wave symmetry is given by%
\begin{equation}
f_{f}^{2\text{D}}\left( \mathbf{k}\right) =k_{x}\left(
k_{x}^{2}-3k_{y}^{2}\right) .
\end{equation}%
The corresponding tight-binding Hamiltonian is given by\cite{GI,Planar,MTJ}%
\begin{equation}
H_{f}=4J\sigma _{z}\sin ak_{x}\sin \frac{ak_{x}+\sqrt{3}ak_{y}}{2}\sin \frac{%
-ak_{x}+\sqrt{3}ak_{y}}{2}.
\end{equation}%
The kinetic term of the tight-binding model on the triangular lattice is
given by 
\begin{equation}
H_{\text{Kine,Tri}}=\frac{\hbar ^{2}}{ma^{2}}\left( 3-\cos ak_{x}-\sum_{\pm
}\cos a\frac{k_{x}\pm \sqrt{3}k_{y}}{2}\right) .  \label{T-tri}
\end{equation}%
The Fermi surface is shown in Fig.\ref{FigF}(a). The mirror symmetry $%
k_{x}\longmapsto -k_{x}$\ is broken, which enables nonzero second-order
nonlinear spin-Seebeck current.

The Seebeck current is calculated as%
\begin{align}
\frac{j_{\text{charge}}^{\left( x;x\right) }}{\frac{e\tau }{\hbar }\partial
_{x}T}=& -\frac{2\beta ^{2}\pi \left( 2-15\pi +m\mu \left( -2+9\pi \right)
\right) }{9\sqrt{3}m^{3}}  \notag \\
& +\frac{2\beta ^{2}\pi \left( J^{2}m^{2}\left( 4+9\pi \left( -2+m\mu
\right) \right) \right) }{9\sqrt{3}m^{3}}.
\end{align}%
The second-order nonlinear spin-Seebeck current is induced as%
\begin{align}
& \frac{j_{\text{spin}}^{(x^{2};x)}}{e\left( \frac{\tau }{\hbar }\partial
_{x}T\right) ^{2}}  \notag \\
=& \frac{\beta ^{3}\pi \left( 14-4m\mu +9\pi +9m\pi \left( -\mu
+6J^{2}m\left( m\mu -2\right) \right) \right) }{3\sqrt{3}m^{3}}sJ.
\end{align}%
It is shown in Fig.\ref{FigF}(b) as well as numerically obtained result
without using the high-temperature expansion.

\begin{figure}[t]
\centerline{\includegraphics[width=0.48\textwidth]{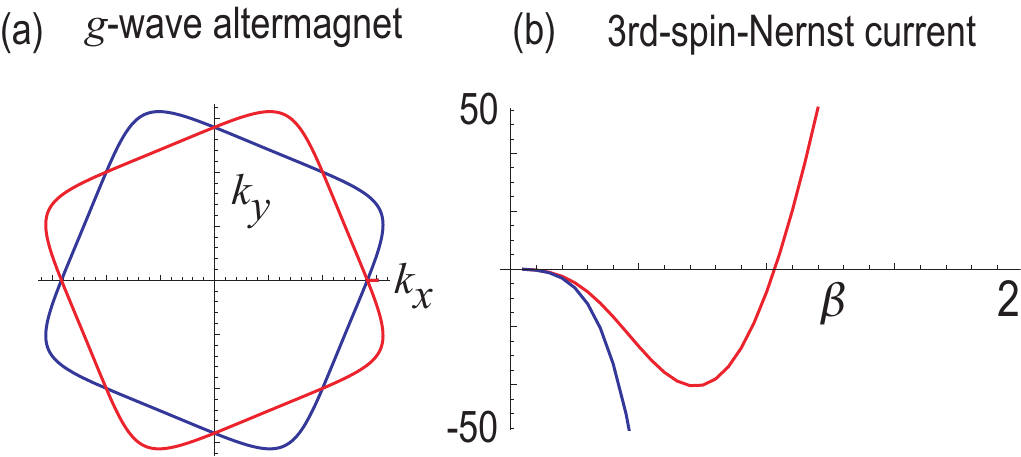}}
\caption{$g$-wave altermagnet. (a) Fermi surfaces with four nodes. (b) Spin
current $j_{\text{spin}}^{\left( x;y\right) }/\partial _{x}T$ in units of $e%
\protect\tau /\hbar $ as a function of the inverse temperature $\protect%
\beta =1/k_{\text{B}}T$. See also the caption of Fig.\protect\ref{FigD}.}
\label{FigG}
\end{figure}

\begin{figure}[t]
\centerline{\includegraphics[width=0.48\textwidth]{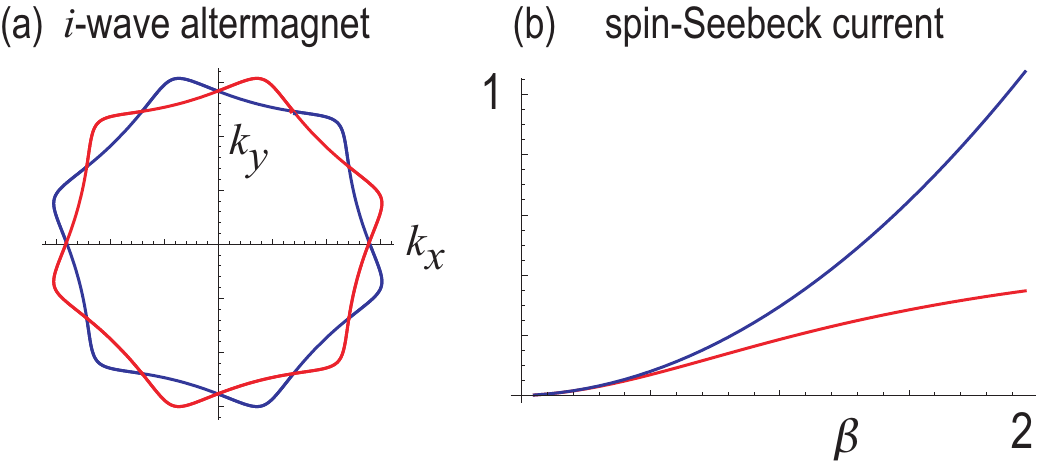}}
\caption{$i$-wave altermagnet. (a) Fermi surfaces with six nodes. (b) Spin
current $j_{\text{spin}}^{\left( x;y\right) }/\left( \partial _{x}T\right)
^{3}$ in units of $e\left( \protect\tau /\hbar \right) ^{3}$ as a function
of the inverse temperature $\protect\beta =1/k_{\text{B}}T$. See also the
caption of Fig.\protect\ref{FigD}.}
\label{FigI}
\end{figure}

$g$\textbf{-wave altermagnet:} The $g$-wave symmetry is given by%
\begin{equation}
f_{g}^{2\text{D}}\left( \mathbf{k}\right) =4k_{x}k_{y}\left(
k_{x}^{2}-k_{y}^{2}\right) .
\end{equation}%
The corresponding tight-binding Hamiltonian is given by\cite{GI,Planar,MTJ}%
\begin{equation}
H_{g}=2J\sigma _{z}\sin ak_{x}\sin ak_{y}\left( \cos ak_{y}-\cos
ak_{z}\right) .
\end{equation}%
The kinetic term of the tight-binding Hamiltonian on the square lattice is
given by%
\begin{equation}
H_{\text{Kine,Sq}}=\frac{\hbar ^{2}}{ma^{2}}\left( 2-\cos ak_{x}-\cos
ak_{y}\right) .  \label{T-sq}
\end{equation}%
The Fermi surface is shown in Fig.\ref{FigG}(a). The mirror symmetry $%
k_{y}\longmapsto -k_{y}$\ is broken, which enables nonzero nonlinear
spin-Nernst current.

The Seebeck current is calculated as

\begin{equation}
\frac{j_{\text{charge}}^{\left( x;x\right) }}{\frac{e\tau }{\hbar }\partial
_{x}T}=-\frac{\beta ^{2}\pi ^{2}\left( 40J^{2}m^{2}+1\right) \left( m\mu
-2\right) }{m^{3}}.
\end{equation}%
The third-order nonlinear spin-Nernst effect is calculated as%
\begin{equation}
\frac{j_{\text{spin}}^{(x^{3};y)}}{e\left( \frac{\tau }{\hbar }\partial
_{x}T\right) ^{3}}=\frac{18\beta ^{4}\pi ^{2}\left( 40J^{2}m^{2}-1\right)
\left( m\mu -2\right) }{m^{4}}sJ.
\end{equation}%
It is shown in Fig.\ref{FigG}(b) as well as numerically obtained result
without using the high-temperature expansion.

$i$\textbf{-wave altermagnet:} The $i$-wave symmetry is given by%
\begin{equation}
f_{i}^{2\text{D}}\left( \mathbf{k}\right) =2k_{x}k_{y}\left(
3k_{x}^{2}-k_{y}^{2}\right) \left( k_{x}^{2}-3k_{y}^{2}\right) .
\end{equation}%
The corresponding tight-binding Hamiltonian is given by\cite{GI,Planar,MTJ}%
\begin{align}
H_{i}=& -\frac{16}{3\sqrt{3}}J\sigma _{z}  \notag \\
& \times \sin ak_{x}\sin \frac{ak_{x}+\sqrt{3}ak_{y}}{2}\sin \frac{-ak_{x}+%
\sqrt{3}ak_{y}}{2}  \notag \\
& \times \sin \sqrt{3}ak_{y}\sin \frac{3ak_{x}+\sqrt{3}ak_{y}}{2}\sin \frac{%
-3ak_{x}+\sqrt{3}ak_{y}}{2}.
\end{align}%
The Fermi surface is shown in Fig.\ref{FigI}(a). The mirror symmetry $%
k_{y}\longmapsto -k_{y}$\ is broken, which enables nonzero nonlinear
spin-Nernst current.

The Seebeck current is calculated as%
\begin{align}
\frac{j_{\text{charge}}^{\left( x;x\right) }}{\frac{e\tau }{\hbar }\partial
_{x}T} =&-\frac{\beta ^{2}\pi }{243\sqrt{3}m^{3}}[\left( 27\left( 2-15\pi
+m\mu \left( 9\pi -2\right) \right) \right)  \notag \\
&+\left( 2J^{2}m^{2}\left( 160+567\pi \left( m\mu -2\right) \right) \right)
].
\end{align}%
The spin-Nernst current is calculated as%
\begin{equation}
\frac{j_{s}^{\left( x;y\right) }}{\frac{e\tau }{\hbar }\partial _{x}T}=\frac{%
-4\beta ^{2}\pi \left( 3m\mu -5\right) }{27\sqrt{3}m^{2}}sJ.
\end{equation}%
It is shown in Fig.\ref{FigI}(b) as well as numerically obtained result
without using the high-temperature expansion.

\textbf{Discussions:} The second-order spin-Seebeck effect $%
j^{(x^{2};x)}\propto \left( \partial _{x}T\right) ^{2}$ occurs only in $f$%
-wave magnets as in Table \ref{TableA}, which is particularly interesting.
There is net spin current even if the temperatures are equal at the right
and left hand sides provided that there is a fluctuation of temperature
because spin current flows regardless of the direction of the temperature
gradient. It acts as a spin current diode.

There is no spin current generation in $X$-wave magnets with $X=p,f,g,i$ in
three dimensions except for $d$-wave altermagnets. Details are shown in
Supplementary Materials II, III and IV.

We have derived analytic formulas for spin currents\ in the leading order of
temperature gradient in a system where the spin is diagonal and hence the
single band is relevant. These formulas are not applicable in the presence
of the spin-orbit interaction such as the Rashba interaction. Current
formulas become complicated because the Berry phase and the quantum metric
are involved for multi-band systems\cite{Vars,Vars2,YFZ}. The order of $J$
in these magnets is 100meV, while the order of the Rashba interaction is
1meV.\ Then, the $J$ term is dominant. Hence, our results will be applicable
even if there is the Rashba interaction.

We mention materialization of $f$-wave magnets, $g$-wave and $i$-wave
altermagnets in two dimensions. An $f$-wave magnet is theoretically proposed
in Ba$_{3}$MnNb$_{2}$O$_{9}$\cite{Hayami2020}, FePO$_{4}$\cite{Hayami2020B}
and in graphene by introducing spin nematic order\cite{BitanRoy}. A $g$-wave
and an $i$-wave altermagnets are theoretically proposed in twisted magnetic
Van der Waals bilayers\cite{YLiu}. An $i$-wave altermagnet is also
theoretically proposed in MnP(S,Se)$_{3}$\cite{MazinIwave}.

The author is very much grateful to M. Hirschberger and E. Saito for helpful
discussions on the subject. This work is supported by CREST, JST (Grants No.
JPMJCR20T2) and Grants-in-Aid for Scientific Research from MEXT KAKENHI
(Grant No. 23H00171).

%%%%%%%%%%%%%%%


\begin{thebibliography}{99}
\bibitem{Dya} M. I. Dyakonov and V. I. Perel, Possibility of orientating
electron spins with current Sov. Phys. JETP Lett. 13: 467, (1971).

\bibitem{Dya2} M. I. Dyakonov and V. I. Perel, Current-induced spin
orientation of electrons in semiconductors. Phys. Lett. A 35 (6): 459 (1971).

\bibitem{Sinova} Jairo Sinova, Sergio O. Valenzuela, J. Wunderlich,
C.\thinspace H. Back, and T. Jungwirth, Spin Hall effects Rev. Mod. Phys.
87, 1213 (2015)

\bibitem{Uchida} K. Uchida, S. Takahashi, K. Harii, J. Ieda, W. Koshibae, K.
Ando, S. Maekawa and E. Saitoh, Observation of the spin Seebeck effect,
Nature volume 455, 778 (2008)

\bibitem{Bauer} Gerrit E. W. Bauer, Eiji Saitoh and Bart J. van Wees Spin
caloritronics Nature Materials 11, 391 (2012)

\bibitem{SqChen} Shu-guang Cheng, Yanxia Xing, Qing-feng Sun, and X. C. Xie,
Spin Nernst effect and Nernst effect in two-dimensional electron systems,
Phys. Rev. B 78, 045302 (2008)

\bibitem{Meyer} S. Meyer, Y.-T. Chen, S. Wimmer, M. Althammer, T. Wimmer, R.
Schlitz, S. Geprags, H. Huebl, D. Kodderitzsch, H. Ebert, G. E. W. Bauer, R.
Gross and S. T. B. Goennenwein Observation of the spin Nernst effect, Nature
Materials 16, 977 (2017)

\bibitem{Gao} Y. Gao, S. A. Yang, and Q. Niu, Field induced positional shift
of Bloch electrons and its dynamical implications, Phys. Rev. Lett. 112,
166601 (2014).

\bibitem{Sodeman} I. Sodemann and L. Fu, Quantum nonlinear Hall effect
induced by Berry curvature dipole in time-reversal invariant materials,
Phys. Rev. Lett. 115, 216806 (2015).

\bibitem{Ideue} T. Ideue, K. Hamamoto, S. Koshikawa, M. Ezawa, S. Shimizu,
Y. Kaneko, Y. Tokura, N. Nagaosa, and Y. Iwasa, Bulk rectification effect in
a polar semiconductor, Nat. Phys. 13, 578 (2017).

\bibitem{HLiu} H. Liu, J. Zhao, Y.-X. Huang, W. Wu, X.-L. Sheng, C. Xiao,
and S. A. Yang, Intrinsic second-order anomalous Hall effect and its
application in compensated antiferromagnets, Phys. Rev. Lett. 127, 277202
(2021).

\bibitem{Michishita} Y. Michishita and N. Nagaosa, Dissipation and geometry
in nonlinear quantum transports of multiband electronic systems, Phys. Rev.
B 106, 125114 (2022).

\bibitem{Watanabe} H. Watanabe and Y. Yanase, Nonlinear electric transport
in odd-parity magnetic multipole systems: Application to Mn-based compounds,
Phys. Rev. Res. 2, 043081 (2020).

\bibitem{CWang} C. Wang, Y. Gao, and D. Xiao, Intrinsic nonlinear Hall
effect in antiferromagnetic tetragonal cumnas, Phys. Rev. Lett. 127, 277201
(2021).

\bibitem{Oiwa} R. Oiwa and H. Kusunose, Systematic analysis method for
nonlinear response tensors, J. Phys. Soc. Jpn. 91, 014701 (2022).

\bibitem{AGao} A. Gao, Y.-F. Liu, J.-X. Qiu, B. Ghosh, T.V. Trevisan, Y.
Onishi, C. Hu, T. Qian, H.-J. Tien, S.-W. Chen et al., Quantum metric
nonlinear Hall effect in a topological antiferromagnetic heterostructure,
Science 381, eadf1506 (2023).

\bibitem{NWang} N. Wang, D. Kaplan, Z. Zhang, T. Holder, N. Cao, A. Wang, X.
Zhou, F. Zhou, Z. Jiang, C. Zhang et al., Quantum metric-induced nonlinear
transport in a topological antiferromagnet, Nature 621, 487 (2023).

\bibitem{KamalDas} Kamal Das, Shibalik Lahiri, Rhonald Burgos Atencia,
Dimitrie Culcer, and Amit Agarwal, Intrinsic nonlinear conductivities
induced by the quantum metric, Phys. Rev. B 108, L201405 (2023).

\bibitem{Kaplan} Daniel Kaplan, Tobias Holder and Binghai Yan, Unification
of Nonlinear Anomalous Hall Effect and Nonreciprocal Magnetoresistance in
Metals by the Quantum Geometry, Phys. Rev. Lett. 132, 026301 (2024).

\bibitem{Ohmic} YuanDong Wang, ZhiFan Zhang, Zhen-Gang Zhu, and Gang Su,
Intrinsic nonlinear Ohmic current, Phys. Rev. B 109, 085419 (2024).

\bibitem{Xiang} Longjun Xiang, Bin Wang, Yadong Wei, Zhenhua Qiao, and Jian
Wang, Linear displacement current solely driven by the quantum metric, Phys.
Rev. B 109, 115121 \ (2024).

\bibitem{ZGong} Zhen-Hao Gong, Z. Z. Du, Hai-Peng Sun, Hai-Zhou Lu, X. C.
Xie, Nonlinear transport theory at the order of quantum metric,
arXiv:2410.04995v2

\bibitem{EzawaMetricC} M. Ezawa, Intrinsic nonlinear conductivity induced by
quantum geometry in altermagnets and measurement of the in-plane Neel
vector, Phys. Rev. B 110, L241405 (2024)

\bibitem{YFang} Yuan Fang, Jennifer Cano, and Sayed Ali Akbar Ghorashi,
Quantum Geometry Induced Nonlinear Transport in Altermagnets, Phys. Rev.
Lett. 133, 106701 (2024).

\bibitem{EzawaPNeel} M. Ezawa, Purely electrical detection of the Neel
vector of p-wave magnets based on linear and nonlinear conductivities, Phys.
Rev. B 112, 125412 (2025)

\bibitem{Yu} Xiao-Qin Yu, Zhen-Gang Zhu, Jhih-Shih You, Tony Low, and Gang
Su, Topological nonlinear anomalous Nernst effect in strained transition
metal dichalcogenides Phys. Rev. B 99, 201410(R) (2019)

\bibitem{Karki} D. B. Karki and Mikhail N. Kiselev, Nonlinear Seebeck effect
of SU(2) Kondo impurity, Phys. Rev. B 100, 125426 (2019)

\bibitem{Zeng} Chuanchang Zeng, Snehasish Nandy, A. Taraphder, and Sumanta
Tewari, Nonlinear Nernst effect in bilayer WTe2 Phys. Rev. B 100, 245102
(2019)

\bibitem{March} G. Marchegiani, A. Braggio, and F. Giazotto, Nonlinear
Thermoelectricity with Electron-Hole Symmetric Systems, Phys. Rev. Lett.
124, 106801 (2020)

\bibitem{Arisawa} Observation of nonlinear thermoelectric effect in
MoGe/Y3Fe5O12 Hiroki Arisawa, Yuto Fujimoto, Takashi Kikkawa and Eiji Saitoh
Nature Communications 15, 6912 (2024)

\bibitem{Hirata} Y. Hirata, T. Kikkawa, H. Arisawa, E. SaitohNonlinear
Seebeck effect in at room temperature, Appl. Phys. Lett. 126, 252408 (2025)

\bibitem{Vars} Harsh Varshney, Amit Agarwal, Intrinsic nonlinear Nernst and
Seebeck effect, New J. Phys. 27, 083506 (2025)

\bibitem{Vars2} Harsh Varshney, Amit Agarwal, Asymmetric Scattering Drives
Large Nonlinear Nernst and Seebeck Effects, arXiv:2601.17775

\bibitem{YFZ} Ying-Fei Zhang, Zhi-Fan Zhang, Hua Jiang, Zhen-Gang Zhu, Gang
Su, Fundamental Relations as the Leading Order in Nonlinear Thermoelectric
Responses with Time-Reversal Symmetry, arXiv:2601.19625

\bibitem{Hamamoto} Keita Hamamoto, Motohiko Ezawa, Kun Woo Kim, Takahiro
Morimoto, and Naoto Nagaosa, Nonlinear spin current generation in
noncentrosymmetric spin-orbit coupled systems, Phys. Rev. B 95, 224430 (2017)

\bibitem{Kameda} Mai Kameda Daichi Hirobe, Shunsuke Daimon, Yuki Shiomi,
Saburo Takahashi, Eiji Saitoh, Microscopic formulation of nonlinear spin
current induced by spin pumping, Journal of Magnetism and Magnetic Materials
476, 459 (2019)

\bibitem{Hayami22B} Satoru Hayami, Megumi Yatsushiro, and Hiroaki Kusunose,
Nonlinear spin Hall effect in PT -symmetric collinear magnets, Phys. Rev. B
106, 024405 (2022)

\bibitem{Hayami24B} Satoru Hayami, Linear and nonlinear spin-current
generation in polar collinear antiferromagnets without relativistic
spin-orbit coupling, Phys. Rev. B 109, 214431 (2024)

\bibitem{BitanRoy} Sanjib Kumar Das, Bitan Roy, From local spin nematicity
to altermagnets: Footprints of band topology, Phys. Rev. B 111, L201102
(2025).

\bibitem{GI} M. Ezawa, Third-order and fifth-order nonlinear spin-current
generation in g-wave and i-wave altermagnets and perfectly nonreciprocal
spin current in f-wave magnets, Phys. Rev. B 111, 125420 (2025).

\bibitem{SmejX} L. Smejkal, J. Sinova, and T. Jungwirth, Beyond Conventional
Ferromagnetism and Antiferromagnetism: A Phase with Nonrelativistic Spin and
Crystal Rotation Symmetry, Phys. Rev. X, 12, 031042 (2022).

\bibitem{SmejX2} Libor \v{S}mejkal, Jairo Sinova, and Tomas Jungwirth,
Emerging Research Landscape of Altermagnetism, Phys. Rev. X 12, 040501
(2022).

\bibitem{Hayami} S. Hayami, Y. Yanagi, and H. Kusunose, Momentum-Dependent
Spin Splitting by Collinear Antiferromagnetic Ordering, J. Phys. Soc. Jpn.
88, 123702 (2019).

\bibitem{pwave} Anna Birk Hellenes, Tomas Jungwirth, Jairo Sinova, Libor 
\v{S}mejkal, Unconventional p-wave magnets, arXiv:2309.01607.

\bibitem{He} T. Jungwirth, R. M. Fernandes, E. Fradkin, A. H. MacDonald, J.
Sinova, L. Smejkal, From supefluid 3He to altermagnets, arXiv:2411.00717

\bibitem{Naka} Makoto Naka, Satoru Hayami, Hiroaki Kusunose, Yuki Yanagi,
Yukitoshi Motome and Hitoshi Seo, Spin current generation in organic
antiferromagnets, Nat. Com. 10, 4305 (2019).

\bibitem{Gonza} Rafael Gonzalez-Hernandez, Libor \v{S}mejkal, Karel Vborn,
Yuta Yahagi, Jairo Sinova, Tom\v{s} Jungwirth, and Jakub \v{Z}elezn,
Efficient electrical spin splitter based on nonrelativistic collinear
antiferromagnetism, Phys. Rev. Lett., 126:127701, (2021).

\bibitem{NakaB} M Naka, Y Motome, and H Seo, Perovskite as a spin current
generator. Phys. Rev. B, 103, 125114, (2021).

\bibitem{Bose} Arnab Bose, Nathaniel J. Schreiber, Rakshit Jain, Ding-Fu
Shao, Hari P. Nair, Jiaxin Sun, Xiyue S. Zhang, David A. Muller, Evgeny Y.
Tsymbal, Darrell G. Schlom \& Daniel C. Ralph, Tilted spin current generated
by the collinear antiferromagnet ruthenium dioxide, Nature Electronics 5,
267 (2022).

\bibitem{NakaRev} Makoto Naka, Yukitoshi Motome and Hitoshi Seo,
Altermagnetic Perovskites, npj Spintronics 3, 1 (2025)

\bibitem{APEX} Motohiko Ezawa, Quantum geometry and X-wave magnets with
X=p,d,f,g,i, arXiv:2512.05477 to be publihsed in APEX.

\bibitem{Planar} M. Ezawa, Almost half-quantized planar Hall effects in
X-wave magnets with X=p,d,f,g,i Phys. Rev. B 112, 235307 (2025)

\bibitem{MTJ} M. Ezawa, Tunneling magnetoresistance in a junction made of
X-wave magnets with X=p,d,f,g,i, arXiv:2509.16867

\bibitem{Hayami2020} S. Hayami, Y. Yanagi, and H. Kusunose, Spontaneous
antisymmetric spin splitting in noncollinear antiferromagnets without
spin-orbit coupling, Phys. Rev. B 101, 220403 (2020)

\bibitem{Hayami2020B} S. Hayami, Y. Yanagi, and H. Kusunose, Bottom-up
design of spin-split and reshaped electronic band structures in
antiferromagnets without spin-orbit coupling: Procedure on the basis of
augmented multipoles, Phys. Rev. B 102, 144441 (2020)

\bibitem{YLiu} Yichen Liu, Junxi Yu, and Cheng-Cheng Liu, Twisted Magnetic
Van der Waals Bilayers: An Ideal Platform for Altermagnetism, Phys. Rev.
Lett. 133, 206702 (2024)

\bibitem{MazinIwave} I. Mazin, R. Gonzalez-Hernandez, and L. Smejkal,
Induced monolayer altermagnetism in mnp(s,se)3 and fese arXiv:2309.02355
\end{thebibliography}
\end{document}